\documentclass[prd,nofootinbib,floats,superscriptaddress,eqsecnum,tightenlines,11pt]{revtex4}
\usepackage{hyperref}
\usepackage{graphicx}
\usepackage{amsmath,amssymb,amsfonts,amsthm,latexsym,stmaryrd}
\usepackage{marginnote}
\usepackage{color}

\def\beq{\begin{equation}}
\def\eeq{\end{equation}}
\newcommand{\bea}{\begin{eqnarray}}
\newcommand{\eea}{\end{eqnarray}}

\def\SU{\text{SU}}

\def\SL{\text{SL}}

\newcommand{\cG}{{\mathcal G}}

\newcommand{\cL}{{\mathcal L}}
\newcommand{\cH}{{\mathcal H}}
\newcommand{\cM}{{\mathcal M}}

\newcommand{\cD}{{\mathcal D}}
\newcommand{\cC}{{\mathcal C}}

\def\SU{\text{SU}}

\def\SL{\text{SL}}

\renewcommand{\(}{\left(}
\renewcommand{\)}{\right)}
\newcommand{\nn}{\nonumber}

\def\be{\begin{equation}}
\def\ee{\end{equation}}
\def\bea{\begin{eqnarray}}
\def\eea{\end{eqnarray}}
\def\nn{\nonumber}

\def\SU{\text{SU}}
\def\SL{\text{SL}}

\begin{document}

\title{Covariance in self dual inhomogeneous models of effective quantum geometry: \\ Spherical symmetry and Gowdy systems}

\author{Jibril Ben Achour}
\affiliation{Center for Field Theory and Particle Physics, 200433 Shanghai, China}
\author{Suddhasattwa Brahma}
\affiliation{Asia Pacific Center for Theoretical Physics, Pohang 37673, Republic of Korea}


\begin{abstract}
When applying the techniques of Loop Quantum Gravity (LQG) to symmetry-reduced gravitational systems, one first regularizes the scalar constraint using holonomy corrections, prior to quantization. In inhomogeneous system, where a residual spatial diffeomorphism symmetry survives, such modification of the gauge generator generating time reparametrization can potentially lead to deformations or anomalies in the modified algebra of first class constraints. When working with self-dual variables, it has already been shown that, for spherically symmetric geometry coupled to a scalar field, the holonomy-modified constraints do not generate any modifications to general covariance, as one faces in the real variables formulation, and can thus accommodate local degrees of freedom in such inhomogeneous models. In this paper, we extend this result to Gowdy cosmologies in the self-dual Ashtekar formulation. Furthermore, we show that the introduction of a $\bar{\mu}-$scheme in midisuperspace models, as is required in the `improved dynamics' of LQG, is possible in the self-dual formalism while being out of reach in the current effective models suing real-valued Ashtekar-Barbero variables. Our results indicate the advantages of using the self-dual variables to obtain a covariant loop regularization prior to quantization in inhomogeneous symmetry reduced polymer models, implementing additionally the crucial $\bar{\mu}$-scheme, and thus a consistent semiclassical limit.
\end{abstract} 

\maketitle

\newpage

\tableofcontents

\newpage

\section{Introduction}

\noindent
Effective models of loop quantum gravity (LQG) provide an interesting platform to study the corrections to General Relativity (GR) induced by loop quantization. The strategy developed in such models is to implement some of the aspects of the loop regularization, prior to quantization, and study the resulting equations at the effective level. The hope is that such an effective treatment will match the quantum dynamics of the system for some well-defined semi-classical quantum states.

The motivation for this strategy lies in its remarkable success when applied to the cosmological scenario. In loop quantum cosmology (LQC) \cite{Ashtekar:2011ni}, the effective dynamics of the universe is obtained by regularizing the extrinsic curvature by a bounded and periodic function, which encodes effectively holonomy corrections. The resulting effective dynamics allows one to capture non-perturbative effects, such as that of singularity resolution. Another form of effective corrections, dubbed triad corrections, are related to the regularization of the inverse-volume term, and lead to additional corrections to the dynamics. Moreover, one can show that such effective dynamics is consistent with the quantum dynamics obtained from semi-classical quantum states (with minimal spread) highlighting the consistency of the effective approach, at the very least, in this homogeneous context \cite{bo1}\footnote{The robustness of the effective Friedman equation has been checked in the simplest case of a massless scalar field minimally coupled to a flat FLRW geometry, but beyond that simple model, it is not clear if higher moments encoding quantum energy fluctuations spoil or not the effective LQC pictures as emphasized in \cite{bo2}}. 

Since then such effective canonical methods have been used in a variety of contexts. The interior black hole geometry, which reduced in static coordinates to an anisotropic cosmological problem, was studied along this line in \cite{Ashtekar:2005qt, Bohmer:2007wi, Corichi:2015xia, Olmedo:2017lvt}. Gravitational systems involving inhomogeneities have also been investigated using the same strategy. The effective corrections to the cosmological perturbations were investigated along this line in \cite{Cailleteau:2011kr, Cailleteau:2012fy}, providing interesting signature of the effective quantum corrections in the CMB power spectra \cite{Linsefors:2012et}. See \cite{Barrau:2014maa, Grain:2016jlq} for reviews. Effective spherically symmetric black hole models were also studied using this effective approach. The anomaly freedom problem in spherical symmetry was worked out in \cite{Tibrewala:2012xb} for inverse-triad corrections (along with an attempt to include a $\bar{\mu}$-scheme) and in \cite{Bojowald:2015zha, Bojowald:2016vlj, Bojowald:2016itl} for holonomy corrections. The anomaly freedom of the (polarized) Gowdy system was treated in \cite{Bojowald:2015sta}. Besides investigating the covariance of these models, effective black holes metrics for the quantum corrected Schwarzschild and Reissner Nordstrom black hole, and for the LTB space-time, were derived in \cite{Tibrewala:2012xb, Bojowald:2009ih, Bojowald:2011js, BenAchour:2018khr, Bojowald:2018xxu}. 

In this article, we would pursue, and extend the investigation initiated in \cite{BenAchour:2016brs}, concerning the self dual formulation of such symmetry reduced effective models. As shown in our previous work, this alternative formulation allows to overcome several difficulties encountered recently in the construction of these inhomogeneous models based on the real Ashtekar-Barbero variables, namely the deformation of general covariance under point-wise holonomy corrections and the difficulty in implementing the so-called $\bar{\mu}$-scheme. Before presenting our results, let us briefly review several crucial aspects of the effective treatment of symmetry reduced canonical models of loop quantum gravity with the real Ashtekar-Barbero variables in order to clarify the motivation of our investigation.

\subsubsection{Partial loop regularization: the choice of effective corrections}
In all models considered in the literature so far, based on the real Ashtekar-Barbero variables, the loop regularization is introduced only partially. For a homogeneous background, the holonomy of the connection reduces to a simple exponentiation, which allows tractable computations. However, when treating inhomogeneous spacetimes, the different components of the connection and the associated electric field inherit different densities depending on their transformation under the residual spatial diffeomorphism. While the connection components having a zero density can be simply exponentiated to treat them as matrix elements of their corresponding holonomies, the components transforming as a density one field involves regularizing extended holonomies. In this case, one can always make such corrections negligible by picking up a small path for the integration. This is not the case for (local) point-wise holonomies. 

This point is best illustrated in the context of the spherically symmetric background. In the real Ashtekar-Barbero formulation, the phase space consists of two canonical pairs (after gauge fixing the Gauss constraint) given by
\be
\{ K_x (x), E^x(y)\} = 2G \delta{(x-y)} \qquad \{ K_{\phi} (x), E^{\phi}(y)\} = G \delta{(x-y)}.
\ee
Under the residual spatial diffeomorphism surviving the symmetry reduction, which corresponds to Lie derivative in the purely radial direction $x$, the components $\left( K_x, K_{\phi}\right)$ do not have the same density since under a coordinate transformation $x \rightarrow \tilde{x}= f(x)$ of the radial coordinate, they behave as
\be
K_{\tilde{x}}(t,\tilde{x}) d\tilde{x} = \frac{\partial \tilde{x}}{\partial x}K_{x}(t,x) dx \qquad \text{while} \qquad K_{\phi}(t,\tilde{x}) d\phi = K_{\phi}(t,x) d\phi
\ee
The former is a density one field while the latter is a zero density field, i.e. a scalar. This has  important consequences when considering their respective holonomies in order to regularize the scalar constraint. The extended holonomy of $K_x$ can be formally computed from a suitable expansion around the value it takes at a given point $x_0$ and turns out to involve higher power of $K_x(x_0)$ as well as higher derivatives, ie $K^{(n)}_x(x_0)$. Indeed, we have for this expansion that
\begin{align}
 h_e (K_x) & = \text{exp} \left( i \gamma \int^{x_0 + \ell_0}_{x_0}  K_x (t,x) dx \right) \nn \\
& = \text{exp} \left( i \gamma \left(  \ell_0 K_x(x_0) + \frac{1}{2}\ell^2_0 K'_x(x_0) + \frac{1}{6} \ell^3_0 K''_x(x_0) + ... \right)  \right) \nn \\
& = 1 + i \gamma \ell_0 K_x(x_0) + \frac{1}{2} \ell^2_0\left( i \gamma K'_x(x_0) - \gamma^2 K^2_x(x_0) \right) + \frac{1}{6} \ell^3_0 \left( i \gamma K''_x (x_0) - 3 \gamma^2 K_x(x_0) K'_x(x_0) - i \gamma^2 K^3_x(x_0)\right)  \nn \\
\label{eh}
& \;\;\; + ... 
\end{align}
and an additional expansion of $\ell_0$, if it is triad dependent, can be performed, leading to an even more complicated expression. A detailed discussion concerning the treatment of such extended holonomy corrections in effective inhomogeneous situation can be found in \cite{Bojowald:2014rma}. 
Naturally, the resulting expression (\ref{eh}) drastically complicates the effective equations, and up till now, no consistent implementations of such corrections have been carried out in any symmetry-reduced model of loop quantum gravity. On the contrary, the holonomy of the zero-density field $K_{\phi}$ can be written as a simple exponentiation, as
\be
h_e(K_{\phi}) = \text{exp} \left(i \int^{\rho}_0  K_{\phi} (t,x) d\phi \right)= \text{exp}\left( i \rho K_{\phi} \right)
\ee
This simple expression allows to introduce this point-wise holonomy corrections within the classical scalar constraint and derive tractable effective field  and look for their solutions.
Moreover, contrary to the extended holonomies of $K_x$, one cannot make this effective correction small by choosing an arbitrarily small parameter $\rho$. Finally note that for homogeneous backgrounds, where all spatial diffeomorphisms are trivial, all the fields transform as zero density field, and this difficulty disappears. 

From this discussion, we can extract two guiding lines which are shared by any effective model of loop quantum gravity when selecting the effective corrections:
\begin{enumerate}
\item Only the zero density connection fields which involve point-wise holonomy corrections are taken into account in the (partial) loop regularization.
\item The effective corrections affecting the density one connection fields are disregarded since they involve higher derivatives terms and since one can make them arbitrary small, contrary to the point-wise corrections.
\end{enumerate}
Note that this strategy is the one shared by any symmetry reduced models of LQG. Because most of the investigations focus on homogeneous models where all the holonomy corrections reduce to point wise corrections, this choice is generally implicit. However, in inhomogeneous models, the difference in the regularization of density zero and density one fields appears explicitly, and the general strategy 1) and 2) is the one adopted so far in the literature.

\subsubsection{The $\bar{\mu}$ versus $\mu_0$-scheme in inhomogeneous effective models}
When introducing effective modifications such as point-wise holonomy corrections, one introduces in the same time a new scale $\rho$. This scale is sometimes called the \textit{polymer scale}. In earlier stage of LQC, this scale was assumed to be a constant, dubbed the $\mu_0$ scheme.  But it was soon realized that such assumption allows the quantum universe to bounce at an arbitrary low energy density and not only close to the Planck one, spoiling therefore the semi-classical consistency of the model. The way out was to assume a metric dependent polymer scale, dubbed the $\bar{\mu}$ scheme, leading to the improved dynamics of LQC \cite{Ashtekar:2006wn}. Hence, the point-wise regularization of the homogeneous and isotropic connection of LQC, i.e. $A^i_a = c(t) \delta^i_a$, was modified as
\begin{align}
\label{mubar}
f_{\mu_0}(c) = \text{exp}\left( i \mu_0 c \right) \qquad \rightarrow \qquad f_{\bar{\mu}}(c, p) = \text{exp}\left( i \bar{\mu} c \right) = \text{exp}\left( i \lambda \frac{c}{\sqrt{p}} \right) \qquad \text{with} \qquad \bar{\mu}(p) = \frac{\lambda}{\sqrt{p}}.
\end{align}
where $p$ is the homogeneous and isotropic component of the electric field $E^a_i = p(t) \delta^a_i$ and $\lambda$ a constant scale. This modification $f(c) \rightarrow f(p,c)$ is crucial for LQC to possess the right semi-classical limit\footnote{Notice that the same modification was recently implemented in the deformed algebra approach to the cosmological perturbations in LQC \cite{Han:2017wmt}.}. The same is true for models of static black hole interior, which are again homogeneous model and in which one can implement the similar techniques as in LQC \cite{Corichi:2015xia}. 

On the contrary, the consequences of working within a $\mu_0$ versus a $\bar{\mu}$ scheme has been much less investigated in inhomogeneous models. Indeed, in most of the investigations dealing with the inhomogeneous spherically symmetric background \cite{Gambini:2013ooa, Gambini:2008dy}, or Gowdy system \cite{Banerjee:2007kh, deBlas:2017goa}, the holonomy corrections are implemented within the $\mu_0$-scheme prior to quantization. While the quantum theory can be constructed, it generates inconsistencies when solving the effective fields equations. This is again best illustrated within the spherically symmetric framework.

The effective field equations for the spherically symmetric static geometry, within the $\mu_0$-scheme, were first investigated in \cite{Bojowald:2009ih, Tibrewala:2012xb} and more recently in \cite{BenAchour:2018khr, Bojowald:2018xxu}. When partially fixing the gauge freedom and asking for a vanishing shift vector $N^x =0$, the resulting set of differential equations implies that the angular component of the extrinsic curvature $K_{\phi}$ is quantized following 
\be
K_{\phi} = \frac{n \pi}{2\rho} \qquad \text{with} \qquad n \in \mathbb{N}^{+}
\ee
Hence, within the $R$-region (outside the horizon), the $\mu_0$-effective corrections imply that the only consistent solution is the Schwarzschild classical solution, corresponding to $n=0$, ie $K_{\phi}=0$. The other solutions are inconsistent since $K_{\phi}$ blows up when $\rho \rightarrow 0$, in the semi-classical limit. In the $T$-region, one can find interesting solutions, but the gluing with the exterior Schwarzschild solution at the horizon fails to be $\cC^1$-differentiable, again because of the $\mu_0$-scheme \cite{Bojowald:2018xxu, BenAchour:2018khr}. Such inconsistencies could be avoided only if the polymer scale $\rho$ becomes metric dependent, which will further change drastically the effective field equations to solve. Hence, the improved regularization performed in LQC seems to be also crucial in spherical symmetry. Such generalization can be written as
\be
\label{barmu}
f_{\mu_0}(K_{\phi}) \qquad \rightarrow \qquad f_{\bar{\mu}}(K_{\phi}) = f_{\lambda}(E^x, K_{\phi})
\ee
where $E^x$ is the only component of the electric field which has zero density and $\lambda$ a constant scale.

As far as we know, the only attempt to generalize the point-wise holonomy corrections according to (\ref{barmu}) in the spherically symmetric case, was performed in \cite{Tibrewala:2012xb}. Unfortunately, it was shown that the correction $f_{\lambda}(E^x, K_{\phi})$ cannot enjoy an interpretation in term of $SU(2)$ holonomy corrections anymore. The reason is that when computing the algebra of the first class modified constraints, the anomaly free conditions imply that the effective correction $f(E^x, K_{\phi})$ cannot be a periodic (or almost-periodic) function anymore. 

Finally, the $\bar{\mu}$-scheme is also crucial when deriving effective actions encoding the point-wise holonomy corrections of loop gravity. Working within the limiting curvature scenario, one could derive scalar-tensor effective actions which reproduces precisely, in the cosmological context, the effective dynamics of LQC implemented in the desired $\bar{\mu}$-scheme \cite{Liu:2017puc, Bodendorfer:2017bjt}. The generalization of this mechanism to spherical symmetry was studied in \cite{BA}. From this investigation, it appears that the $\bar{\mu}$-scheme is a necessary (although not sufficient) ingredient to derive a covariant effective action encoding the point-wise holonomy corrections.

From this discussion, we summarize two important lessons concerning the implementation of the $\bar{\mu}$-scheme in effective inhomogeneous models:
\begin{enumerate}
\item The current regularization of the spherically symmetric background done in the $\mu_0$-scheme exhibits inconsistencies when solving the effective field equations. Just as in LQC, one needs to introduce an improved dynamics based on a regularization implementing a $\bar{\mu}$-scheme, i.e. metric dependent point-wise holonomy corrections.
\item  The first attempt to implement a $\bar{\mu}$-scheme in inhomogeneous spherically symmetric background, using real Ashtekar-Barbero variables, shows that the resulting effective corrections can no longer be matrix elements of $SU(2)$ holonomies, if they satisfy the anomaly-free condition, thereby presenting us with a major impediment.
\end{enumerate}
On the one hand, the modification (\ref{barmu}) is necessary for the consistency of an effective treatment, but represents a non trivial step which has not yet been successfully implemented in inhomogeneous effective models of (real-valued) LQG.

\subsubsection{Anomaly freedom in inhomogeneous effective models}
A natural question which arises when working out such effective models concerns the fate of covariance in the underlying gravitational systems. Indeed, when introducing holonomy or inverse-triad corrections to the classical scalar constraint, one does modify the generator of time reparametrization, which is a first class constraint. Hence, the effective modifications introduced by the (partial) loop regularization can, a priori, spoil the gauge symmetry of the underlying gravitational system, which can potentially lead to the appearance of spurious degrees of freedom. Therefore, a careful analysis of the anomaly problem is required in order to ensure that the effective modifications introduced in the classical first class constraints do not spoil the gauge symmetry of the system. In a canonical formulation, this is achieved by computing Dirac's hypersurface deformation algebra (DHDA) of the modified constraints. In the full theory, this algebra takes the well known classical form
\begin{align}
& \{ D[U^a], D[V^b]\} = D[\cL_{{U}} V^a] \\
&  \{ D[U^a], \cH[N]\} =  - \cH[\cL_{{U}} N]   \\
& \{ \cH[N], \cH[M]\} =  D[ q^{ab} \big{(} N \partial_b M - M \partial_b N\big{)}]  \, , \label{Ham}
\end{align}
where we have used the notations: $U=U^a \partial_a$ and $V=V^a\partial_a$ are spatial vectors, $N$ and $M$ are scalars and ${\cL}_{{U}}$ denotes the usual Lie derivative along the vector $U$. $q^{ab}$ is the inverse spatial metric and $D[U^a]$ and $H[N]$ are the smeared vectorial and scalar constraints respectively.

In the minisuperspace cosmological context, the problem of anomalies disappears because only the scalar constraint survives the symmetry reduction to homogeneous background, and the DHDA is trivial. Therefore, effective cosmological models are blind to such potential covariance issues. However, when going beyond homogeneous models, such as for the treatment of cosmological perturbations, or a spherically symmetric background, anomaly-freedom becomes crucial for the consistency of the effective approach. Let us now briefly summarize the results obtained so far for such inhomogeneous gravitational systems.

Anomaly freedom for the vacuum spherically symmetric model was studied in a series of works \cite{Tibrewala:2012xb, Bojowald:2015zha, Bojowald:2015sta, Bojowald:2016vlj, Bojowald:2016itl}. The triad and holonomy corrections were studied separately and both regularizations were shown to generate a deformation of general covariance. Moreover, the holonomy corrections were investigated  both in the old $\mu_0$-scheme (used so far the loop quantization of spherically symmetric inhomogeneous background \cite{Gambini:2013ooa}) as well as in the $\bar{\mu}$-scheme. In this latter improved dynamics, the Dirac's algebra is again deformed such that
\begin{align}
\label{def}
 \{ \cH[N], \cH[M]\} & =  D_{\text{grav}}[ \beta(E^x, K_{\phi}) q^{xx} \big{(} N M' - M N'\big{)}],
\end{align}
while the anomaly-free condition for the two holonomy corrections $f_1(E^x, K_{\phi})$ and $f_2(E^x, K_{\phi})$ read
\begin{align}
\label{AFC}
f_2 - \frac{1}{2} \frac{\partial f^2_1}{\partial K_{\phi}}  + 2 E^x \frac{\partial f_1}{\partial E^x} = 0,
\end{align}
where $\beta = \partial f_2 / \partial K_{\phi}$. It turns out that the condition (\ref{AFC}) spoils the possibility to implement periodic (or almost periodic) functions as holonomy corrections since choosing  $f_2(E^x, K_{\phi})$ such that
\begin{align}
f_2 (E^x, K_{\phi}) = \left( E^x\right)^p \sin\left( \left( \frac{1}{E^x}\right)^p K_{\phi} \right)
\end{align}
one obtains for $f_1(E^x, K_{\phi})$ that
\begin{eqnarray}
f_{1,p} &=& \sqrt{2} \left(- \left( E^x\right)^p \left((1+ 4p) \left( E^x\right)^p \cos\left( \left( \frac{1}{E^x}\right)^p K_{\phi} \right) \right)\right.\nn\\
&& \left.+ 2 p K_{\phi} \sin\left( \left( \frac{1}{E^x}\right)^p K_{\phi} \right)  + (1+4p) \left( E^x\right)^{2p} \right)^{1/2},
\end{eqnarray}
choosing the last term such that the classical limit is consistent, i.e. $f_{1,p} \rightarrow K_{\phi}$ when $1/ E^x \rightarrow 0$. See \cite{Tibrewala:2012xb} for details.

Thus, the price of anomaly freedom, which has been largely ignored so far, implies that the implementation of a $\bar{\mu}$-scheme, which is required for semi-classical consistency, is still out of reach of such models. This no-go result call thus for a generalization of the loop regularization used so far in such inhomogeneous models of LQG.

Similar investigations were performed for the inhomogeneous Gowdy system with rotational symmetry as well as two dimensional dilatonic gravity, the CGHS black hole model bing a special case of the latter \cite{Bojowald:2016vlj}. A general proof of the deformation of the general covariance due to the holonomy corrections (within the $\mu_0$ scheme) in such midi-superspace model was presented in \cite{Bojowald:2016itl}. Hence, deformed covariance appears to be a generic property of the current loop regularization where only point-wise holonomy corrections can be consistently implemented. Moreover, all these results concern symmetry-reduced gravitational system having no local degrees of freedom.

The introduction of any local degrees of freedom turn out to be even more problematic within the current approach. In the context of spherically symmetric background, local degrees of freedom can be introduced via matter couplings. In general, the gravitational and matter sectors do not have a matching notion of covariance\footnote{Note that the Einstein-Maxwell system is an exception of this case, and the Dirac's algebra is deformed but closed after regularization. However, it also does not have any local degrees of freedom.}. It results in an anomalous DHDA of the modified constraints such that
\begin{align}
 \{ \cH[N], \cH[M]\} & =  D_{\text{grav}}[ \beta(K_{\phi}) q^{xx} \big{(} N M' - M N'\big{)}] + D_{\text{matter}}[ q^{xx} \big{(} N M' - M N'\big{)}]  \, , \label{Ham} \\
 & \neq D_{\text{full}}[ \beta(K_{\phi}) q^{xx} \big{(} N M' - M N'\big{)}]
\end{align}
and prevent any quantization of the loop regularized system \cite{Bojowald:2015zha}. Specifically, this class of no-go results prevent the development of a model of gravitational collapse based on realistic matter field, and with it, the phenomenon of Hawking radiation. The case of spherical symmetry could suggest that the problem arises because of the coupling to matter, but such interpretation would be misplaced. In the case of the vacuum Gowdy system without local rotational symmetry, where gravitational waves are present, one faces the same anomaly problem as in spherical symmetry \cite{Bojowald:2015sta}. Hence, the problem is intimately related to the presence of any local degrees of freedom, and occurs even in vacuum configuration.

In the present work, we will show that the above issues, namely
\begin{itemize}
\item the consistent implementation of holonomy corrections in the $\bar{\mu}$-scheme as required for semi-classical consistency, and 
\item the anomaly showing up in the Dirac's algebra of holonomy-modified first-class constraints in presence of local degrees of freedom
\end{itemize}
can both be overcome by using the self dual Ashtekar variables. Since the above no-go results derived within the real variables formulation concern either the inverse-triad or the holonomy corrections separately, we will focus on the second partial loop regularization involving only the holonomy corrections. By applying precisely similar regularizations, we will show how the no-go results of \cite{Bojowald:2015zha} can be by-passed. But before presenting the computation, let us justify why working with the self dual variables is a valid strategy when considering the off-shell anomaly freedom in midi-superspace models.

\subsubsection{Why use self-dual Ashtekar variables?}
The original motivation for working with the self dual variables comes mainly from a set of intriguing results obtained over the last five years in black hole thermodynamics.
As it is well known, the state counting procedure based on the $\SU(2)$ Chern-Simons theory for spherically symmetric isolated horizon, based on the real Ashtekar-Barbero variables, successfully led to the derivation of the Bekenstein-Hawking area law \cite{Ashtekar:2000eq, Engle:2010kt}. Yet, this result requires a self tuning of the real Barbero-Immirzi parameter $\gamma$ \footnote{The fine tuning of the Barbero-Immirzi parameter in the spherically symmetric case led to an alternative model for the entropy state counting \cite{Ghosh:2011fc, Ghosh:2013iwa, Asin:2014gta}. Based on this model, the dependecy on this parameter of the leading term is shifted to the sub-leading terms for the spherically symmetric horizon. Yet, the axisymmetric case remains problematic within the standard techniques. Moreover, the above alternative relies, among other ingredients, on the assumption that the degeneracy of the horizon depends exponentially on the area.}. Moreover, the axisymmetric case was treated in two different models, leading to two rather different results for the state counting \cite{Ashtekar:2001is, Frodden:2012en}. Within the framework of \cite{Frodden:2012en}, it was shown that the standard state counting is not consistent since one cannot recover the expected area law, whatever the value of the BI parameter is \cite{BenAchour:2016mnn}, in contrast with the conclusion obtained in \cite{Ashtekar:2004nd}. 

In a series of papers \cite{Frodden:2012dq, BenAchour:2016mnn, Achour:2014eqa, Achour:2015xga}, it was shown that one can consistently analytically continue the dimension of the Hilbert space of the isolated horizon from $\gamma \in \mathbb{R}$ to $\gamma = \pm i$, both in the spherically symmetric and axisymmetric case. The resulting entropy turns out to match exactly the expected semi-classical result for both types of horizons. Moreover, the degeneracy of the horizon depends exponentially on the area after the analytic continuation, contrary to the case $\gamma \in \mathbb{R}$ where it depends on the area through a power law. It was also shown in \cite{Geiller:2014eza} how the self dual formulation allows to recover naturally the thermal character of the near-horizon radiation. Additional results have been obtained in this context, reinforcing the status of the self dual variables as well suited to capture the right semi-classical result in the context of black hole thermodynamics in LQG \cite{Pranzetti:2013lma, Pranzetti:2014tla, Frodden:2012nu, Heidmann:2016yfz}. See also \cite{Geiller:2012dd, Achour:2013hga, Achour:2013gga, Achour:2014rja} for related investigations in different contexts.

From the perspective of the symmetry carried by the self dual formulation, it is also well known fact that within the parameter space of LQG, the point $\gamma = \pm i$ enjoys special properties. The self dual connection is indeed a true space-time connection, which transforms properly under the scalar constraint, contrary to its real counterpart. It is then natural to expect difficulties when implementing the dynamics (of symmetry reduced models) within the real Ashtekar-Barbero formulation. This problem indeed re-emerges in effective inhomogeneous polymer models where the loop regularization generically deformed the Dirac's algebra. When the scalar constraint is regularized using holonomy corrections, the Lorentzian term $\cH^{L} \propto (1 + \gamma^2) E^a_i E^b_j  K^{[i}_a K^{j]}_b/ \sqrt{\text{det}(E)}$ involves second order spatial derivatives of the triad. This is precisely this term which generates the deformation $\beta{(K_{\phi})}$ in the bracket (\ref{def}). 
Since this term disappears from the scalar constraint when $\gamma = \pm i$, the self dual formulation is not affected by this problem and no deformation appears, as shown in \cite{BenAchour:2016brs}\footnote{See also \cite{BenAchour:2016leo} for the inhomogeneous cosmological case.}. In what follow, we extend this result by showing that the self dual formulation also allows to implement a covariant notion of holonomy corrections within the $\bar{\mu}$-scheme. We stress that this point is not a trivial generalization of the result already obtained in \cite{BenAchour:2016brs} for spherical symmetric geometry. As already explained in detail above, the implementation of a $\bar{\mu}$-scheme in inhomogeneous models, crucial for the consistency of the approach, is out of reach within the current techniques as shown in \cite{Tibrewala:2012xb}. Therefore, the result obtained in this work further underlines the peculiar status of the self dual formulation in the context of inhomogeneous models.
 
Let us conclude this introduction by a brief comment on reality conditions. It is well-known that this is a serious problem when dealing with the self dual formulation in full gravity. However, in symmetry-reduced models such as spherically symmetric gravity, where the Dirac observables are known from the beginning, one can still work with self dual variables and impose reality conditions on the level of the physical Hilbert space by turning the operators corresponding to Dirac observables self-adjoint. This strategy was used in earlier investigations on the canonical quantization of self dual spherically symmetric vacuum and electro-vacuum gravity in \cite{Thiemann:1992jj, Thiemann:1994qb}. This is the strategy we wish to pursue in future work when quantizing the full model corresponding to the holonomy-corrected constraints which we have studied in this work.

 \section{Spherically symmetric self dual gravity with an effective loop regularization}

\noindent The goal of this section is to extend the work done in \cite{BenAchour:2016brs}  it by including holonomy modifications in the so-called $\bar{\mu}-$scheme, as would be pertinent for the `improved-dynamics' formulation of LQG.

\subsection{Spherically symmetric self dual Ashtekar gravity}
\noindent The symmetry reduction of the self dual phase space to spherically symmetric geometry has been worked out in details in \cite{Thiemann:1992jj}. The spherically symmetric background metric reads
\be
ds^2 = - (N^2 - N^xN_x) dt^2 + q_{xx} dx^2 + 2 N^x dx dt + q_{\theta\theta} d\Omega^2
\ee
such that
\be
q_{xx} = \frac{E}{2E^1} \qquad \text{and} \qquad q_{\theta\theta} = E^1
\ee
where $E = (E^2)^2 + (E^3)^2$ is the Gauss invariant quantity built from the angular components of the electric field. Within this symmetry-reduced framework, the action reduces to
\be
S [A_1, A_2, A_3, E^1, E^2, E^3, \Lambda, N, N^x] = \frac{4\pi}{\kappa} \int_{\cM} dt dx \; \big{[} - i E^i \dot{A}_i - \Phi \big{]} +  \frac{4\pi}{\kappa} \int_{\partial \cM} dt  \; \text{Boundary}
\ee
where the factor $4\pi$ comes from the integration of the action over the angular coordinates and the Lagrange multipliers $\Lambda, N, N^x$ appear in addition to the canonical pairs $(A_i, E^i)$. All the canonical variables do not have the same density: $A_1, E^2, E^3$ are density-one fields while $E^1, A_2, A_3$ are scalars. This will be relevant when introducing holonomy corrections. The term $\Phi$ represents the linear combination of the first class constraints while the last term `$\text{B\,dy}$' stands for the boundary term due to integrating by part. Therefore, it contains the charges of the geometry under different symmetries of the canonical variables. Explicitly,
\be
\Phi = i \Lambda \cG - i N^x \cH_x + \frac{1}{2} \tilde{N} \cH 
\ee
while 
\begin{align}
\text{Boundary} = - i \Lambda E^1 + i N^x (A_2 E^2 + (A_3 - \sqrt{2})) E^3 ) + \tilde{N} E^1 (A_2 E^3 - (A_3 - \sqrt{2}) E^2)\,.
\end{align}
Note that we work with a densitized lapse $\tilde{N} = N/ \sqrt{q}$. The first class constraints are explicitly given by
\begin{align}
\label{Gauss}
& \cG = (E^1)' + A_2 E^2 - A_3 E^3\,, \\
\label{SpatialDiffeo}
& \cH_x  = B^2 E^3 - B^3 E^2\,, \\
\label{Scalar}
& \cH  = \frac{1}{2} \; \big{[} \; E^2 (2 B^2 E^1 + B^1E^2) + E^3 (2 B^3 E^1 + B^1 E^3) \; \big{]}\,,
\end{align}
and one denotes the spatial diffeomorphism constraint as
\be
\cD_x = \cH_x - A_1 \cG = - A_1 (E^1)' +(A_2)' E^2 + (A_3)' E^3\,.
\ee
The variable $B^i = B^i_a dx^a$ are dual to the curvature two form,  $B^{i \; c} = \frac{1}{2} \epsilon^{abc} F^i_{ab}$. Its components are given by
\begin{align}
\label{B1}
B^1 & = \frac{1}{2} \; \big{[}\; (A_2)^2 + (A_3)^2 - 2 \; \big{]}\,, \\
B^2 & = A'_3 + A_1 A_3\,, \\
B^3 & = - A'_2 + A_1 A_3\,.
\end{align}
The full Hamiltonian is fully constrained as a result
\begin{align}
H_{\text{full}}  = \frac{4\pi}{\kappa}  \int_{\Sigma} dr \; \big{\{}  i \Lambda \cG - i N^x \cH_x + \frac{1}{2} \tilde{N} \cH  \;\big{\}}  -  \frac{4\pi}{\kappa} \int_{\partial \cM} dt  \; \text{B\,dy}
\end{align}
This conclude our presentation of the classical self dual phase space of spherically symmetric gravity.
\subsection{The effective quantum corrections}\label{Sec}
\noindent We would like to investigate the fate of general covariance of the holonomy-corrected spherically symmetric geometry. The effective quantum corrections one can introduce are restricted because, thus far, one does not have full control on extended holonomies. Therefore, the density one fields, such as $A_1$, cannot be regularized easily following usual loop techniques. This problem has been discussed in much of the literature, say \cite{Brahma:2014gca}. In order to incorporate such extended holonomies, one should introduce non local functions, which is still an open problem in inhomogenous models of LQG \cite{Bojowald:2014rma}. Therefore, as a first pass, one considers only pointwise holonomies, affecting only the scalar components of the connection\footnote{This makes the quantum-corrected constraint ultra-local, which can lead to some problems in the infra-red (see, for instance, \cite{deBlas:2017goa}.)}. A formal description of such point-wise modifications should naturally read
\be
A_2 \rightarrow f_{\mu}(A_2) \qquad \text{and} \qquad A_3 \rightarrow f_{\mu}(A_3)
\ee
where $\mu$ is a constant length/energy scale on which the effective quantum correction depends. 

However, working in the self dual formulation, one can write expressions for the first class constraints, on the phase space, purely in term of the dual of the curvature, $B^i$. Since this dual formulation is fully determined by the components of the self dual connection, and its derivatives, one can work with the variable $B^i$ instead of $A^i$, both being equivalent classically. Since it is unclear how one can regularize the inhomogenous componenent $A_1$, one can only regularize the components of dual curvature $B^i$ which does not involve $A_1$. The only component which we can regularize is therefore $B_1 \sim (A_2)^2 + (A_3)^2 - 2 $. (Working with the self-dual connection, as opposed the extrinsic curvature components to base the holonomy corrections, it can be shown that these are the most general corrections which one can introduce \cite{UsFuture}.) The above-mentioned effective corrections can, hence, be written as
\be
B^1 \rightarrow f_{\mu}(B^1)\,.
\ee
This is the effective correction which was considered in \cite{BenAchour:2016brs}. However, such corrections do not take into account the so called $\bar{\mu}-$scheme, which states that the polymer scale $\mu$ one introduces during the regularization should be dynamical, depending explicitly on the geometry, through the radial electric field component $E^1$. This is required since a constant length scale would not only spoil the diffeomorphism symmetry, but also a dynamical scale is required to recover the proper classical limit and ensure that LQG corrections do not spoil the GR solution. In most midisuperspace LQG models, this $\bar{\mu}-$scheme is not taken into account when using the real Ashtekar-Barbero variables. The reason is that, albeit the quantum-corrected constraints form a closed algebra even in the $\bar{\mu}-$scheme, the effective correction functions are necessarily of a form such that they cannot be (quasi-)periodic anymore \cite{Tibrewala:2012xb}. It then becomes challenging to interpret them as  holonomy corrections of some compact gauge group, as is required from the fundamental theory. 

We will show in the next two sections that the self dual formulation allows one to include this improved regularization scheme in an anomaly-free version of the quantum-corrected system and without putting any prohibitive restrictions on the holonomy-correction functions that would spoil their interpretations as $\SL(2,\mathbb{C})$ holonomies. Moreover, one obtains an undeformed notion of covariance, without modifying the structure functions as is the case for the real-variables. In the following, we shall denote the improved $\bar{\mu}-$scheme effective correction as
\be
\label{CorHOL}
B^1 \qquad \rightarrow \qquad f_{\bar{\mu}} (B^1) = f_{\mu} (E^1, B^1)\,.
\ee
Having discussed the general form of our effective quantum corrections, we can now turn to the expression of the holonomy-corrected first class constraints. As is usual in polymer models, we do not modify the momentum constraint nor the Gauss constraint. The reason is that the (spatial) diffeomorphism symmetry is usually implemented through a group averaging procedure, applied to the gauge invariant spin-network states. In this process, one obtains diffeomorphism invariant quantum states of the geometry, without resorting to the infinitesimal expression of the quantum  diffeomorphism constraint. Only the scalar constraint, when imposed to the diff-invariant states, is regularized using holonomy and fluxes. (In any case, one does not expect the quantum geometry corrections to affect the internal gauge symmetry of the system, thereby leaving the Gauss constraint unmodified.) The Hamiltonian constraint is therefore modified in the following manner
\begin{align}
 \cH  = \frac{1}{2} \; \big{[} \; E^2 (2 B^2 E^1 + f_{\mu}(E^1, B^1) E^2) + E^3 (2 B^3 E^1 + g_{\mu}(E^1, B^1) E^3) \; \big{]}\,,
\end{align}
while the unmodified Gauss, vector and spatial diffeomorphism constraints read
\begin{align}
& \cG  = (E^1)' + A_2 E^2 - A_3 E^3\,, \\
& \cH_x  = B^2 E^3 - B^3 E^2\,, \\
& \cD_x = - A_1 (E^1)' +(A_2)' E^2 + (A_3)' E^3\,.
\end{align}

In the following section, we shall investigate the fate of the Dirac algebra for the above holonomy-corrected constraints. We will show that the effective corrections introduced still allow for a closed algebra, which ensures that no spurious unphysical degrees of freedom propagate. But even more, as we shall see, the self dual LQG model turns out to have the same Dirac hypersurface deformation algebra as its classical counterpart, ensuring the same notion of general covariance as in GR.

\subsection{Algebra of the holonomy-corrected constraints: the gravitational sector}
\noindent
In order to investigate the fate of the covariance, and therefore the closure and probable deformation of the Dirac algebra, we shall compute the brackets between the first-class constraints $\cD_x$ and $\cH$. In the following, we will omit the tilde ($\sim$) on the densitized lapse to simplify notation.
\subsubsection{The $\{H[N], H[M]\}$ bracket}
\noindent
This is the bracket in which the inverse of spatial metric appears as a structure function and thus, the usual deformation of the Dirac algebra appears in this one for the real variables. Let us reproduce of the Hamiltonian constraint, smeared with the lapse function, in the presence of modification functions for the benefit of the reader
\begin{eqnarray}
\label{Scal}
 H[N] = \frac{1}{2} \int d x\; N \left[E^2 (2 B^2 E^1 + f_{\mu}(E^1, B^1) E^2) + E^3 (2 B^3 E^1 + g_{\mu}(E^1, B^1) E^3)\right]\,.
\end{eqnarray}
It is useful to recall that $B^2, B^3$ depend on spatial derivatives whereas $B^1$ is independent of them since the only brackets between conjugate variables, which do not cancel each other, are the ones in which one of the terms have a spatial derivative. If we keep this in mind, then following the calculations of \cite{BenAchour:2016brs}, we can immediately see that making the functions $f$ and $g$ dependent on $E^1$ does not affect the structure of the algebra. This is so because including any functional dependence on $E^1$ does not affect any of the brackets since none of $(B^1, B^2, B^3)$ depend on the spatial derivative of $A_1$. Thus making the holonomy modifications dependent on $E^1$ does not change the outcome of the bracket. Thus, we have the same structure of the bracket between two Hamiltonian constraints as in the classical case even when we introduce $\bar{\mu}-$dependent holonomy modifications. 

Let us next show this by explicit calculation. The bracket between the first and third terms is given by
\bea\label{TermA}
& &\int \text{d}x \text{d}y M(x)N(y) E^1(x)E^1(y)\left\{E^2(x)B^2(x), E^3(y)B^3(y)\right\} - (x \leftrightarrow y)\nn\\
&=& i\int \text{d}x\, \text{d}y M(x)N(y) E^1(x)E^1(y)\left(E^2(x)B^3(y) \frac{\text{d}}{\text{d}x}\left[\delta(x,y)\right] + B^2(x)E^3(y) \frac{\text{d}}{\text{d}y}\left[\delta(x,y)\right]\right) - (x \leftrightarrow y) \nn\\
&=& i\int \text{d}x \left(M'(x)N(x)-N'(x)M(x)\right) \left(\left(E^1(x)\right)^2 E^2(x) B^3(x) - \left(E^1(x)\right)^2 E^3(x) B^2(x)\right)\,.
\eea
Since the terms involved in the above bracket does not have the presence of any modification functions and is exactly identical to the classical calculation. Looking at terms involving modification functions, we calculate the bracket between the first and fourth terms
\bea\label{TermB}
& &\int \text{d}x \text{d}y M(x)N(y) E^1(x)E^2(x)\,g_\mu\left(E^1(y), B^1(y)\right)\left\{B^2(x), \left(E^3(y)\right)^2\right\} - (x \leftrightarrow y)\nn\\
&=&\int \text{d}x \text{d}y M(x)N(y) E^1(x)E^2(x)\, g_\mu\left(E^1(y), B^1(y)\right)E^3(y) \frac{\text{d}}{\text{d}x}\left[\delta(x,y)\right] - (x \leftrightarrow y)\nn\\
&=& i\int \text{d}x \left(M'(x)N(x)-N'(x)M(x)\right) E^1(x)E^2(x)E^3(x)\,g_\mu\left(E^1(x), B^1(x)\right)\,.
\eea
Proceeding similarly, we find that the bracket between the second and third terms gives
\bea\label{TermC}
-i\int \text{d}x \left(M'(x)N(x)-N'(x)M(x)\right) E^1(x)E^2(x)E^3(x)\,f_\mu\left(E^1(x), B^1(x)\right)\,.
\eea
By requiring that $\left\{H[N], H[M]\right\}$ close into the vector constraint,  as in the classical case, we have
\bea
\left[H[N], H[M]\right] = i\int \text{d}x \left(M(x)N'(x)-N(x)M'(x)\right) \left(E^1(x)\right)^2 \left\{B^2(x)E^3(x) - B^3(x)E^2(x)\right\}\,,
\eea
where the vector constraint is given by $H^x[N^x] = \int \text{d}x N^x(x)\left\{B^2(x)E^3(x) - B^3(x)E^2(x)\right\}$.
It is immediately obvious that we require $f_\mu\left(E^1, B^1\right) = g_\mu\left(E^1, B^1\right)$ for (\ref{TermB}) to cancel (\ref{TermC}), in order to have $\left\{H[N], H[M]\right\}$ brackets close into the vector constraint, just as in the classical case.

\subsubsection{The $\{ H[N], D[N^x]\}$ bracket}
\noindent
Once again, it is helpful to recall the calculation of this bracket for the $\mu_0$ scheme as was done in \cite{BenAchour:2016brs} in order to extend it for the $\bar{\mu}-$scheme. Rewriting the diffeomorphism constraint
\be
\label{Vec}
D[N^x]=\int \text{d}x N^x [-A_1(E^1)^\prime + A_2^\prime E^2 + A_3^\prime E^3]\,,
\ee
we want to evaluate the bracket $\left\{D[N^x], H[N]\right\}$. In this case, it is sufficient to only examine of the holonomy-modified terms from the Hamiltonian constraint since its bracket with the diffeomorphism constraint must reproduce that specific term. Thus, if the bracket of one such modified term can be reproduced successfully, then it would work for all other terms as well.  (The brackets between terms which are left unmodified in the Hamiltonian constraint, and the diffeomorphism constraint, obviously has the same form as in the classical calculation.) We choose the bracket between the diffeomorphism constraint and the second term in the Hamiltonian constraint to illustrate our claims.
\bea
&&\left\{D[N^x], -\frac{1}{2}\int \text{d}y N f_\mu(B^1, E^1)(E^2)^2\right\}\nn\\
&=& \frac{1}{2}\int\text{d}x\text{d}y N^x(x)N(y) \left[-A_1(x)(E^1)^\prime(x)+ A_2(x)^\prime E^2(x)+A_3^\prime(x) E^3(x), f_\mu(B^1(y), E^1(y))(E^2)^2(y)\right]\nn\\
&=& -i\int\text{d}x N^{x\prime} N f_\mu(B^1, E^1) (E^2)^2 -i\int \text{d}x N^xN f_\mu(B^1, E^1) (E^2)^\prime E^2\nn\\
& &\,\,\, -\frac{i}{2}\int \text{d}x N^xN A_2^\prime \frac{\partial f_\mu}{\partial A_2}(E^2)^2 -\frac{i}{2}\int \text{d}x N^xN A_3^\prime \frac{\partial f_\mu}{\partial A_3} (E^2)^2 -\frac{i}{2}\int \text{d}x N^xN (E^1)^\prime \frac{\partial f_\mu}{\partial E^1} (E^2)^2\nn\\
&=& -\frac{i}{2}\int \text{d}x \left(N^{x\prime} N - N^x N'\right) (E^2)^2 f_1(B^1, E^1) -\frac{i}{2}\int \text{d}x N^{x\prime} N (E^2)^2 f_\mu(B^1, E^1) \nn\\
& &\,\,-\frac{i}{2}\int \text{d}x  N^x N'(E^2)^2 f_\mu(B^1, E^1) -\frac{i}{2}\int \text{d}x N^x N (E^2)^2 [f_\mu(B^1, E^1)]^\prime -i\int \text{d}x N^xN f_\mu(B^1, E^1) (E^2)^\prime E^2\nn\\
&=& -\frac{i}{2}\int \text{d}x \left((N^x)^\prime N - N^x N'\right) (E^2)^2 f_\mu(B^1, E^1) -\frac{i}{2}\int \text{d}x (N^xN)^\prime (E^2)^2 f_1(B^1, E^1)\nn\\
& &\,\,\, -\frac{i}{2}\int \text{d}x N^xN (E^2)^2 [f_\mu(B^1, E^1)]^\prime -\frac{i}{2}\int \text{d}x N^xN f_\mu(B^1, E^1)\left((E^2)^2\right)^\prime\nn\\
&=& -\frac{i}{2}\int \text{d}x \left((N^x)^\prime N - N^x N'\right) (E^2)^2 f_\mu(B^1, E^1) + \text{total derivative}\,.
\eea
In the above, we have used the relation 
\be
f_\mu' =
(\partial f_\mu / \partial A_2) A_2' + (\partial f_\mu / \partial A_3) A_3' + (\partial f_\mu / \partial E^1) (E^1)'
\ee
As we have shown, this bracket remains unmodified as in the classical case
\bea
\left\{D[N^x],H[N]\right\} = H[\cL_{N^x} N]
\eea

\subsubsection{The $\{ D[N^x_1], D[N^x_2]\}$ bracket}
\noindent
Obviously the $\{D,D\}$ bracket remains the same since we do not put any holonomy corrections in the (spatial) diffeomorphism constraint.
Therefore, we conclude that the point-wise holonomy corrections (\ref{CorHOL}) introduced above, in the $\bar{\mu}-$scheme, do not modify the Dirac hypersurface deformation algebra and the modified effective system remains covariant in the sense of GR.

\subsection{Adding a scalar field with an arbitrary potential}
\noindent
So far, we have only considered the pure gravitational contribution to the constraints. In order to study properties such as Hawking radiation in LQG models of the Schwarzschild black hole, we need to add matter and study the ensuing system. Therefore, looking ahead, we choose to work with a scalar field with an arbitrary potential, whose contribution to the Hamiltonian constraint reads
\bea
H_{\text{scalar}}=\int \text{d}^3x \,N\, \left\{  \frac{P_\Phi^2}{2\sqrt{q}} - \frac{1}{2}\sqrt{q}q^{xx}\Phi'^2 +\sqrt{q} V(\Phi)  \right\}\,,
\eea
whereas the diffeomorphism constraint in this case is
\bea
D_{\text{scalar}}=\int \text{d}^3x\, N^x\, \Phi' P_\Phi \,.
\eea
We have one more canonical pair, given by $\{ P_\Phi(x), \Phi(y) \} = \delta(x,y)/4\pi$, in the (gravity+matter) spherically-symmetric phase space. Thus the full constraints are as follow
\bea
D_T[M^x] &=& D_{\text{grav}}[M^x] + D_{\text{scalar}}[M^x]\nn\\
&=&  \int dx \; M(x) \; ( \; A'_3 E^3 + A'_2 E^2 - A_1 (E^1)'  \;) + 4 \pi \int dx \; M(x) \;  P_{\Phi} \Phi ' \,,\\
H_T[N] &=& H_{\text{grav}}[N] + H_{\text{scalar}}[N] \nn\\
&=&  \frac{1}{2} \int dx \; N(x) \; \; ( \; E^2 ( 2 E^1 B^2 + E^2 f_{\mu}(E^1, B^1) ) + E^3 ( 2 E^1 B^3 + E^3 g_{\mu} (E^1, B^1) ) \; )\nn \\
 & &\,\,\, +\,2 \pi \int dx \; N(x) \; \left( \; P^2_{\Phi} + (E^1)^2 \; \Phi'^{2} + E^1((E^2)^2+(E^3)^2)  \; V(\Phi) \; \right)\,.
\eea

Let us investigate the brackets between these constraints term by term. We have already computed the brackets coming from the pure gravitational sector in the previous section. Therefore, we only have to consider the bracket of the pure matter sector as well as the cross term between the two sectors. Since we didn't introduce any modifications to the matter sector, it is straightforward to see that the brackets $\{H_{\text{scalar}}[N] , H_{\text{scalar}}[M] \}$, $\{H_{\text{scalar}}[N] , D_{\text{scalar}}[M^x] \}$ and $D_{\text{scalar}}[M^x] \}, D_{\text{scalar}}[N^x] \}$ keep their classical form.

It turns out that the cross term will also remain the same. Consider first the bracket $\{ H_{\text{grav}}[N], H_{\text{scalar}}[M]\} + (N \leftrightarrow M)$. Due to the antisymmetry of the Poisson bracket, only terms involving integration by part, and thus having a spatial derivative will contribute. Now, note that the matter contribution to the scalar constraint does not depend on any components of the self dual connection and therefore, only the bracket between the $E^2$ and $E^3$ from the matter sector with the $A_2$ and $A_3$ of the gravitational one could generate undesired terms. However, $A_2$ and $A_3$ only appears in the gravitational contribution through the combination \eqref{B1} which does not appear in the second and fourth terms in the gravitational Hamiltonian \eqref{Scal}. Hence, the contribution to the cross-terms involving these two terms vanishes. The only non-zero bracket is between the first and third terms of the gravitational part of the Hamiltonian constraint with the matter part of the Hamiltonian constraint. But the first and third terms of the gravitational part are precisely those which are left unmodified by holonomy corrections and hence, the cross term between the gravitational and matter parts is cancelled exactly as they were in the classical calculation. 

\section{The self dual (unpolarized) Gowdy system with an effective loop regularization}
\noindent 
In the previous section, we have reviewed and extended the results obtained in \cite{BenAchour:2016brs} for the spherically symmetric geometry minimally coupled to a scalar field with an arbitrary potential. We would like now to consider a new system: the unpolarized Gowdy midisuperspace cosmology. It self dual formulation was first investigated in \cite{Husain:1989qq}. This symmetry-reduced (pure) gravity model have local degrees of freedom corresponding to gravitational waves often called Einstein-Rosen waves. It represents the natural next step when considering more general symmetry reduced gravity models. In the following, we show that the results obtained in the first section extend to this more complicated system.

\subsection{The classical self dual Gowdy system}
\noindent
The phase space of Gowdy cosmology is given by the five canonical pairs of variables, the first one denoted $(A^3_{\theta}, E^{\theta}_3)$ corresponding to the inhomogenous degrees of freedom while the four other ones, $(A^i_{x}, E^{x}_i)$ and $(A^i_{y}, E^{y}_i)$ being the radial coordinates. The canonical Poisson brackets read
\begin{align}
\label{CV}
\{ A^3_{\theta}(\theta), E^{\theta}_3(\theta ')\} = i\kappa \delta(\theta - \theta')\qquad \{ A^i_{a}(\theta), E^{b}_j(\theta ')\} = i\kappa \delta^i_j \delta^b_a \delta(\theta - \theta')
\end{align}
where the indice $a$ runs over the two indices $(x,y)$ and the index $3$ corresponds to the third internal directions. Since none of the field variables depend on the the $x$ or $y$ coordinate, it is possible to integrate over them (physically, this would mean integrating over the 2-torus). Following the calculations of \cite{Husain:1989qq}, and similar calculations for real-valued variables in \cite{Banerjee:2007kh}, and  we have imposed a gauge fixing, which is given by the following, in order to reduce the phase space.
\begin{align}
& E^x_3 = E^y_3 = E^{\theta}_2 = E^{\theta}_1 = 0 \\
& A^3_x = A^3_y = A^1_{\theta} = A^2_{\theta} = 0
\end{align}
Under this gauge fixing, the constraints of self dual Ashtekar gravity for the Gowdy cosmology reduces to
\begin{align}
& \cG_3 = \partial_{\theta} E^{\theta}_{3} + \epsilon_{IJ} A^I_{a} E^{a\; J} \\
& \cH_{\theta} = E^{a}_I F^I_{\theta a}   \\
& \cH = 2 \epsilon_{IJ} \; E^{\theta}_3 E^{a \; I} F^J_{\theta a} + \epsilon^{IJ} E^a_I E^b_J F^3_{ab}
\end{align}
where $(I,J)$ are internal indices running now over $\{ 1,2\}$. One can now count the physical degrees for freedom per phase space point. The five canonically conjugate pairs correspond to ten degrees of freedom in total. Since we have three first class constraint surviving the symmetry reduction and the gauge fixing, we end up with $n = 10 - 2 \times 3 = 4$ physical degrees of freedom, meaning that we recover the two configuration degrees of freedom per point in this system. These degrees of freedom correspond to the gravitational waves present in the vacuum Gowdy system. While the variables $(E^\theta_3, A^i_a)$ transform as scalars under diffeomorphisms, the $(E^a_i, A^3_\theta)$ variables transform as density-one objects. 

This gauge fixing corresponding to the Gowdy cosmology naturally obeys the DHDA satisfied by the first class constraints
\begin{align}
& \{\cD[N^{\theta}],  \cD [M^{\theta}]\} = \cD [\cL_{N^{\theta}} M^{\theta}] \\
&\{\cH[N],  \cD[M^{\theta}]\} = - \cH [\cL_{M^{\theta}} N]\\
&\{\cH[N],  \cH[M]\} = D [q^{\theta\theta} (N \partial_{\theta} M - M \partial_{\theta} N )]
\end{align}
where $q^{\theta\theta} = \({E^\theta_3}\)/\(\sqrt{(E^x_1)^2 + (E^x_2)^2} \sqrt{(E^y_1)^2 + (E^y_2)^2}\)$.

For our calculations, it will be useful to have the explicit expressions of the previous constraints in terms of the components of the self dual connections and their dual electric fields. For this purpose, one can introduce the magnetic field $B^{a}_i$, dual to the curvature of the connection, i.e. $B^{i \; c} = \frac{1}{2} \epsilon^{abc} F^i_{ab}$, as was done in the spherically symmetric case. Its components are given by
\begin{align}
& B^{\theta}_3 = A^1_x A^2_y - A^2_x A^1_y  \\
& B^x_1 = - (\partial_{\theta} A^1_y + A^3_{\theta} A^2_y) \\
& B^x_2 = - (\partial_{\theta} A^2_y - A^3_{\theta} A^2_y)  \\
& B^y_1 = (\partial_{\theta} A^1_x + A^3_{\theta} A^2_x)  \\
& B^y_2 = (\partial_{\theta} A^2_x - A^3_{\theta} A^1_x)
\end{align}
We can now rewrite the expression of the first class constraints in term of the canonical variables (\ref{CV}) and the above magnetic field components
\begin{align}
\cG_3 & = \partial_{\theta} E^{\theta}_{3} + E^{x\; 1} A^2_x + E^{y\; 1} A^2_y - E^{x\; 2} A^1_x - E^{y\; 2} A^1_y \\
\cH_{\theta} & = E^x_1 B^y_1 + E^x_2 B^y_2 - E^y_1 B^x_1 - E^y_2 B^x_2 \\
\cH  & = 2 E^{\theta}_3 \; \{  E^{x\; 1} B^y_2 - E^{y\; 1} B^x_2 - E^{x\; 2} B^y_1 + E^{y\; 2} B^x_1 \} + (E^x_1 E^y_2 - E^x_2 E^y_1) B^{\theta}_3
\end{align}

Except for the Gauss constraint, the constraints can be written only with the electric and magnetic field. Finally, the generator of spatial diffeomorphisms reads
\begin{align}
\cD_{\theta} & = A^3_{\theta} \cG_3 - \cH_{\theta}  = A^3_{\theta} \partial_{\theta} E^{\theta}_3 - E^x_1 \partial_{\theta} A^{1}_{x} - E^y_1 \partial_{\theta} A^{1}_{y} - E^x_2 \partial_{\theta} A^{2}_{x} - E^y_2 \partial_{\theta} A^{2}_{y}\,.
\end{align}
Being interested in the algebra of these constraints, we shall consider their smeared versions given by
\begin{align}
D_{\theta}[N^{\theta}] = \frac{1}{i \kappa} \int d\theta \; N^{\theta} \cD_{\theta}\,, \qquad H[N] = \frac{1}{2\kappa} \int d\theta \; N \cH\,.
\end{align}
Having presented the classical self dual phase space of the Gowdy system, we can now discuss the form of the effective holonomy corrections we want to introduce. This is the subject of the next section.

\subsection{The effective quantum corrections}
\noindent
Following what we have done in Sec. \eqref{Sec}, we introduce our quantum corrections at the level of magnetic field. Just as for the spherically symmetric case treated above, we only modify the component which have a zero density, which corresponds to $B^{\theta}_3$. Moreover, we consider the general case where the corrections additionally depend on the homogeneous component of the electric field $E^{\theta}_3$, implementing an effective $\bar{\mu}-$scheme. Hence, the general form of our corrections is of the form
\be
\label{HCorr}
B^{\theta}_3 \qquad \rightarrow \qquad f_{\bar{\mu}}(B^{\theta}_3) = f_{\mu}(E^{\theta}_3, B^{\theta}_3)\,,
\ee
where $\mu$ corresponds to the standard polymer scale. With this general form of the effective quantum corrections, the scalar constraint becomes
\begin{align}
\cH^{\text{qc}} = 2 E^{\theta}_3 \; \{ E^{x\; 1} B^y_2 - E^{y\; 1} B^x_2 - E^{x\; 2} B^y_1 + E^{y\; 2} B^x_1 \} + (E^x_1 E^y_2 - E^x_2 E^y_1) f_\mu(B^{\theta}_3, E^{\theta}_3)\,,
\end{align}
while the generator of the spatial diffeomorphisms remains unmodified
\begin{align}
& \cD_{\theta}  = A^3_{\theta}\;  \partial_{\theta} E^{\theta}_3 - E^x_1 \partial_{\theta} A^{1}_{x} - E^y_1 \partial_{\theta} A^{1}_{y} - E^x_2 \partial_{\theta} A^{2}_{x} - E^y_2 \partial_{\theta} A^{2}_{y}\,.
\end{align}
An additional thing to notice here is that the function $f_\mu(B^{\theta}_3, E^{\theta}_3)$ needs to be an odd function of the argument $B^\theta_3$. Having the general expression for our effective quantum-corrected constraints, we can now look at their algebra and check whether they remain first-class.

\subsection{Algebra of the holonomy-corrected first class constraints}
\noindent
In this section, we investigate the fate of the DHDA for the quantum-corrected constraints on the effective Gowdy phase space.

\subsubsection{The $\{H[N], H[M]\}$ bracket}
\noindent
Since the deformation of the DHDA generically occurs for the bracket of the scalar constraint with itself in the real Ashtekar-Barbero variables, we compute this bracket first. For practical purpose, we split the scalar constraint into two pieces $\cH = \cH_1 + \cH_2$ given by 
\begin{align}
\cH_1 & =  2 E^{\theta}_3 \; \{  E^{x\; 1} B^y_2 - E^{y\; 1} B^x_2 - E^{x\; 2} B^y_1 + E^{y\; 2} B^x_1 \} \\
\cH_2 & =  (E^x_1 E^y_2 - E^x_2 E^y_1) f_\mu(B^{\theta}_3, E^{\theta}_3)
\end{align}
Notice that only the term $\cH_2$ receives effective quantum corrections through $f_\mu\(B^{\theta}_3, E^{\theta}_3\)$. It is then straightforward to see that (by the antisymmetry of the Poisson bracket), $\{ \cH_2[N], \cH_2[M]\}$ vanishes trivially since the term $\cH_2$ doesn't involve derivatives and therefore there can be no terms surviving the integration by parts. 

From these observations, the only bracket we need to compute is the cross term which can potentially generates undesired terms. It reads
\begin{align}
& \{ H_1[N], H_2 [M]\} + \{ H_2[N], H_1 [M]\} \nn \\
& = \frac{1}{2\kappa^2} \int d\theta d\theta'  \;NM\; \left\{ \;  E^{\theta}_3 (E^{x\; 1} B^y_2 - E^{y\; 1} B^x_2 - E^{x\; 2} B^y_1 + E^{y\; 2} B^x_1) \;,\;  (E^x_1 E^y_2 - E^x_2 E^y_1) f_\mu(B^{\theta}_3, E^{\theta}_3) \; \right\} \nn \\
& \qquad -  (N-M)\nn \\
& = \frac{i}{2\kappa} \int d\theta \; (M\partial_{\theta} N - N \partial_{\theta} M) E^{\theta}_3  \big{[} \; E^{x\;1} E^y_1 - E^{y1} E^x_1 - E^{x\; 2} E^y_2 + E^{y\; 2} E^x_2 \; \big{]} f_\mu(B^{\theta}_3, E^{\theta}_3) \nn \\
& = 0
\end{align}
Hence, this cross term vanishes without requiring any conditions of the effective quantum corrections. Moreover, the full bracket reduces to $\{H_1[N], H_1[M]\}$ which keeps its classical form since $\cH_1$ is not affected by the holonomy corrections introduced in \eqref{HCorr}. Thus, we obtain that
\begin{align}
\{H[N], H[M]\} & = H_{\theta} [q^{\theta \theta} (N \partial_{\theta} M - M \partial_{\theta} N)]\,,
\end{align}
which implies that the bracket of the holonomy-corrected Hamiltonian constraint, with itself, keeps its classical expression. 

\subsubsection{The $\{H[N], D_{\theta}[N^{\theta}]\}$ bracket}
\noindent
We turn now to the second bracket between the Hamiltonian constraint and the generator of spatial diffeomorphisms. Using again the splitting of the Hamiltonian constraint, as we introduced above to proceed to the computation, we first consider the bracket
\begin{align}
\label{AVDB}
\{H_1[N], D_{\theta}[N^{\theta}]\} & = \frac{1}{i \kappa^2} \int d\theta d\theta' \; \Bigg\{ N E^{\theta \; 3} \left( E^{x\; 1} B^y_2 - E^{y\; 1} B^x_2 - E^{x\; 2} B^y_1 + E^{y\; 2} B^x_1 \right) ;  \\
& \qquad \qquad \qquad \qquad N^{\theta} \left(A^3_{\theta} \; \partial_{\theta} E^{\theta}_3 - E^x_1 \partial_{\theta} A^{1}_{x} - E^y_1 \partial_{\theta} A^{1}_{y} - E^x_2 \partial_{\theta} A^{2}_{x} - E^y_2 \partial_{\theta} A^{2}_{y} \; \right) \; \Bigg\} \nn 
\end{align}
In the following, we only compute explicitly the bracket involving the first term of $\cH_1$ since the three other are similar and their expression do not bring additional information. This first bracket reads explicitly
\begin{align}
& \frac{1}{i \kappa^2} \int d\theta d\theta' \; \{ N E^{\theta \; 3} E^{x\; 1} B^y_2 \; ; N^{\theta} \; \big{[}A^3_{\theta} \partial_{\theta} E^{\theta}_3 - E^x_1 \partial_{\theta} A^{1}_{x} - E^y_1 \partial_{\theta} A^{1}_{y} - E^x_2 \partial_{\theta} A^{2}_{x} - E^y_2 \partial_{\theta} A^{2}_{y} \; \big{]} \; \} \nn \\
 & = \frac{1}{i \kappa^2} \int d\theta d\theta' \; \{ N E^{\theta \; 3} E^{x\; 1} (\partial_{\theta} A^2_x + A^3_{\theta} A^1_x) \; ; N^{\theta} \; \big{[}A^3_{\theta} \partial_{\theta} E^{\theta}_3 - E^x_1 \partial_{\theta} A^{1}_{x} - E^y_1 \partial_{\theta} A^{1}_{y} - E^x_2 \partial_{\theta} A^{2}_{x} - E^y_2 \partial_{\theta} A^{2}_{y} \; \big{]} \; \} \nn \\
& =   \frac{1}{\kappa} \int d\theta \; \big{[} \; -  (N N^{\theta}) E^{x\; 1} B^y_2 \partial_{\theta} E^{\theta}_3   - N \partial_{\theta} ( N^{\theta} E^x_1) E^{\theta}_3 B^y_2 + N^{\theta} \partial_{\theta} (N E^{\theta}_3 E^{x\; 1}) \partial_{\theta} A^2_x \nn \\
& \qquad \qquad \qquad \qquad \qquad \qquad \qquad \qquad- N \partial_{\theta} (N^{\theta} A^3_{\theta}) E^{\theta}_3 E^x_1 A^1_x - (N N^{\theta}) E^{\theta}_3 E^x_1 A^3_{\theta} \partial_{\theta} A^1_x \; \big{]} \nn \\
& =  \frac{1}{\kappa} \int d\theta \; \big{[} \; (N^{\theta} \partial_{\theta} N)  E^{\theta}_3 E^x_1 (B^y_2 - A^3_{\theta} A^1_x)  - (N \partial_{\theta} N^{\theta}) E^{\theta \; 3} E^{x\; 1} ( B^y_2 + A^3_{\theta} A^2_x) \nn \\
& \qquad \qquad  \qquad  + (N N^{\theta}) ( \; \partial_{\theta} (E^{\theta \; 3} E^{x\; 1}) (B^y_2 - A^3_{\theta} A^2_x)  - \partial_{\theta} (E^{\theta \; 3} E^{x\; 1}) B^y_2 - E^{\theta\; 3} E^{x\; 1} \partial_{\theta} (A^1_x A^3_{\theta})\; ) \; \big{]} \nn \\
& =  \frac{1}{\kappa} \int d\theta \; \big{[} \; (N^{\theta} \partial_{\theta} N)  E^{\theta}_3 E^x_1 \partial_{\theta} A^2_x - (N \partial_{\theta} N^{\theta}) E^{\theta \; 3} E^{x\; 1}  ( B^y_2 + A^3_{\theta} A^2_x)   -   (N N^{\theta})  \partial_{\theta}( E^{\theta \; 3} E^{x\; 1}  A^3_{\theta} A^1_x  ) \; \big{]} \nn \\
\end{align}
Gathering all the terms, the bracket (\ref{AVDB}) becomes
\begin{align}
 \{ H_1 [N], D[N^{\theta}]\} & = \frac{1}{\kappa} \int d\theta \; (N^{\theta} \partial_{\theta} N - N \partial_{\theta} N^{\theta}) \; \big{[}E^{\theta}_3 (- E^x_1 B^y_2 + E^x_2 B^y_1 - E^y_1 B^x_2 + E^y_2 B^x_1) \big{]}  \nn \\
& \;\; + \frac{1}{\kappa} \int d\theta \; (N^{\theta} \partial_{\theta} N - N \partial_{\theta} N^{\theta}) \; \big{[} \; - E^x_1 A^3_{\theta} A^2_x + E^x_2 A^3_{\theta} A^1_x + E^y_1 A^3_{\theta} A^1_y - E^y_2 A^3_{\theta} A^2_y\; \big{]} \nn \\
& \;\; + \frac{1}{\kappa} \int d\theta \; (N^{\theta}  N ) \partial_{\theta} (\; - E^x_1 A^3_{\theta} A^2_x + E^x_2 A^3_{\theta} A^1_x + E^y_1 A^3_{\theta} A^1_y - E^y_2 A^3_{\theta} A^2_y\; ) \nn \\
\label{LastTerm1}
& = - H_1[N^{\theta} \partial_{\theta} N - N \partial_{\theta} N^{\theta}]  + \text{surface term}
\end{align}
where this calculation exactly follows the classical case as this computation does not involve the holonomy-correction introduced in \eqref{HCorr}.

We can now compute the last bracket. Keeping in mind that $f_\mu(E^{\theta}_3, B^{\theta}_3) = f_\mu(E^{\theta}_3, A^1_x A^2_y - A^2_x A^1_y)$, we have that
\begin{align}
&\{H_2[N], D_{\theta} [N^{\theta}] \}\nn \\ 
& =  \frac{1}{2 i \kappa^2} \int d\theta d \theta' \; \big\{ \;  N (E^x_1 E^y_2 - E^x_2 E^y_1) \; f_\mu(E^{\theta}_3, B^{\theta}_3)  ; \nn \\
& \qquad \qquad \qquad \qquad  \qquad \qquad \qquad \qquad \qquad  N^{\theta} \big{[}  A^3_{\theta}\; \partial_{\theta}E_3^{\theta} - E^x_1 \partial_{\theta} A^{1}_{x} - E^y_1 \partial_{\theta} A^{1}_{y}  - E^x_2 \partial_{\theta} A^{2}_{x} - E^y_2 \partial_{\theta} A^{2}_{y} \big{]} \; \big\} \nn \\
& = -\frac{1}{2 \kappa} \int d\theta \; \big{[}\; N \partial_{\theta} (N^{\theta} E^y_2 E^x_1) f_\mu + N \partial_{\theta} (N^{\theta}E^x_1E^y_2) f_\mu - N (\partial_{\theta} N^{\theta} E^y_1 E^x_2) f_\mu - N( \partial_{\theta} N^{\theta} E^x_2E^y_1) f_\mu  \nn \\
& \qquad - ( N  N^{\theta}) (E^x_1 E^y_2 - E^x_2 E^y_1) \; \big{(} \;   \frac{\partial f_\mu}{\partial E^{\theta}_3} \partial_{\theta} E^{\theta}_3  + \frac{\partial f_\mu}{\partial A^1_x} \partial_{\theta} A^1_x + \frac{\partial f_\mu}{\partial A^1_y} \partial_{\theta} A^1_y + \frac{\partial f_\mu}{\partial A^2_x} \partial_{\theta} A^2_x +  \frac{\partial f_\mu}{\partial A^2_y} \partial_{\theta} A^2_y \; \big{)} \; \big{]} \nn \\
& =  -\frac{1}{2\kappa} \int d\theta \;   \big{[} \; 2  (N \partial_{\theta} N^{\theta})\;  (E^x_1 E^y_2 - E^x_2 E^y_1) \; f_\mu(E^{\theta}_3, B^{\theta}_3) \nn \\ 
& \qquad \qquad  \qquad \qquad \qquad \qquad \qquad + N N^{\theta} \big{(} \;  \partial_{\theta} (E^x_1 E^y_2 - E^x_2 E^y_1) \; f_\mu(E^{\theta}_3, B^{\theta}_3) +  (E^x_1 E^y_2 - E^x_2 E^y_1)  \partial_{\theta} f_\mu(E^{\theta}_3, B^{\theta}_3) \; \big{)} \; \big{]} \nn \\
& = -\frac{1}{2\kappa} \int d\theta \; (N^{\theta} \partial_{\theta} N - N \partial_{\theta} N^{\theta})  (E^x_1 E^y_2 - E^x_2 E^y_1) f_\mu(E^{\theta}_3, B^{\theta}_3) + \text{surface term}\nn\\
& = - H_2[N^{\theta} \partial_{\theta} N - N \partial_{\theta} N^{\theta}] + \text{surface term}\label{Lastterm2}
\end{align}
This shows that the second part of the Hamiltonian constraint, which is modified by the holonomy correction function, transforms in the same way under the spatial diffeomorphism generator as in the classical computation.

Putting together \eqref{LastTerm1} and \eqref{Lastterm2}, we find 
\begin{align}
\{H[N], D_{\theta}[N^{\theta}]\} = - H [ (N^{\theta} \partial_{\theta} N - N \partial_{\theta} N^{\theta}) ]\,.
\end{align}
Once again, we recover the classical bracket between the Hamiltonian constraint and the generator of spatial diffeomorphisms even after introducing holonomy corrections in the $\bar{\mu}-$scheme. 

\subsubsection{The $\{ D[N^{\theta}_1], D[N^{\theta}_2]\}$ bracket}
\noindent
Just as in the case of spherical symmetry, the $\{D,D\}$ bracket remains the same since we do not put any holonomy corrections in the (spatial) diffeomorphism constraint.

\section{Discussion}
Symmetry-reduced models provide us with a test-bed for applying quantization techniques of LQG since it is yet unknown how to do so for full $(3+1)$-dimensional gravity. However, in order to model realistic physical scenarios, one has to go beyond homogeneous toy models. On the other hand, physical scenarios such as that of black hole evaporation need coupling to matter degrees of freedom, which are forbidden in models of real Ashtekar-Barbero variables. Moreover, even for vacuum spherically symmetric gravity, holonomy-corrections could not be implemented in the $\bar{\mu}-$scheme, which is what is usually done while obtaining singularity-free models in loop quantum cosmology. Even when one includes holonomy corrections with a fixed length scale, deformed covariance appears generically in LQG models with real variables.

In this work, we managed to achieve both -- include holonomy corrections in the $\bar{\mu}-$scheme as well as preserve an undeformed algebra of (quantum-corrected) constraints, as in GR. This was possible by trading in the real-valued Ashtekar-Barbero variables for their self dual counterparts. This forms the first steps of loop quantizing an inhomogeneous model with local physical degrees of freedom. This is a necessary step in examining the formation of black holes and the dynamical resolution of the classical singularity inside their cores. In addition, we also established similar conclusions for Gowdy cosmologies working within the same formulation. These models have gravitational waves and provide an ideal arena for quantization of such local degrees of freedom. An immediate physical consequence of using the $\bar{\mu}$-scheme in the Gowdy model can be seen as follows. It has been well established that plane-polarized gravitational waves can be studied in LQG, based on the Gowdy model, having imposed further constraints on the phase space \cite{Hinterleitner:2010zx,Hinterleitner:2011rb,Girelli:2012ju}. Indeed, a consistent loop quantization of this system would be a first LQG model of gravity waves (see \cite{Hinterleitner:2017ard} for further developments in this direction). Without a $\bar{\mu}$-scheme, such a quantum theory is bound to be plagued with the appearance of spurious effects such as a bounded curvature in a sub-Planckian regime. This is because the polymer scale in these models is a free-parameter and can be easily fine-tuned, depending on auxiliary structures. As argued by us throughout this paper, one needs to find an implementation of the `improved dynamics' scheme in LQG and our algebraic results forms a first step towards this direction. The use of a $\bar{\mu}$-scheme can, perhaps, also ameliorate some of the problems arising in a model of (polarized) Gowdy cosmology, with an additional rotational symmetry, with regards to Dirac observables in the semi-classical limit \cite{deBlas:2015gha,deBlas:2017goa}.

The significance of our work is twofold. On one hand, we show that the effective-symmetries of these LQG models does not have to be deformed with respect to the classical ones if one uses self-dual variables. In this case, one finds an anomaly-free algebra of quantum-corrected constraints which are also undeformed with respect to the structure-functions. This naturally enables one to bypass the no-go conclusions of \cite{Bojowald:2015zha,Bojowald:2015sta}, while using real-valued variables, and allows for adding matter degrees of freedom to the spherically-symmetric system or even lay down the first step for a loop quantization of the Gowdy model. On the other hand, it also identifies the self-dual connection, and not the extrinsic curvature components, as the suitable variables to base the holonomy corrections on. It is important to note that our results are only valid for the self-dual connection and not just any generic complex-valued connection. Generically, using any connections other than the self dual ones would immediately lead one to find deformations in the LQG-corrected theory. Thus, the absence of deformations point towards a protected symmetry in the theory, pointing towards the self dual variables as a special point in the parameter space. A further line of inquiry would naturally involve exploring why the symmetries of the self dual variables, as generated by the $\SL(2,\mathbb{C})$ gauge group, are protected in the full theory. Finally, as has been noted earlier in related work \cite{BenAchour:2016brs}, our results are not a contradiction to the uniqueness theorem due to Hojman, Kucha\v{r} and Teitelboim \cite{Hojman:1976vp}. Indeed, their result applies to full $(3+1)-$dimensional gravity while ours is for symmetry-reduced models. 


\end{document}